\documentclass[showpacs,showkeys,preprint]{revtex4}

\usepackage{graphicx}

\newlength{\imgwidth}
\begin{document} 

\title{Three Bead Rotating Chain model shows universality in the stretching of proteins}
\author{Hamed Seyed-allaei}
\affiliation{International School for Advanced Studies (SISSA), via Beirut 4, 34014 Trieste , Italy}
\affiliation{INFM, Trieste, Italy}
\email{allaei@sissa.it}
\preprint{66/2004/FSB}

\begin{abstract}

We introduce a model of proteins in which all of the key atoms in the protein backbone are accounted for, thus extending the Freely Rotating Chain model. We use average bond lengths and average angles from the Protein Databank as input parameters, leaving the number of residues as a single variable. The model is used to study the stretching of proteins in the entropic regime. The results of our Monte Carlo simulations are found to agree well with experimental data, suggesting that the force extension plot is universal and does not depend on the side chains or primary structure of proteins.

\end{abstract}
\pacs{87.14.Ee, 87.15.Aa, 87.15.La, 82.37.Rs}
\keywords{ Modeling and Simulation of Proteins, Single Molecule Experiments}

\maketitle
\newpage


\section{Introduction}

Accurate modeling of protein structure and dynamics is an immense challenge in the biological sciences, and has elicited intense interest among physicists. Many models have been developed, from first principle studies to coarsegrained phenomenological theories. A major issue is the tradeoff between inclusion of microscopic details, and computational tractability and efficiency. It is hoped that the latter can be improved by removing select microscopic details which do not affect the accuracy of the results.

In this work we introduce a model of the protein backbone and apply it to the problem of protein stretching, for which many experimental results are available. 
This allow us to model the backbone of proteins accurately without paying attention to effective interactions or force-fields.


\subsection{Experiments}  
We know the elastic property of single molecules by 
atomic force microscopy (AFM) and optical tweezers \cite{gaub,optic1,optic2,non_machanical,fernandez_titin}. 
One of the most studied proteins by these experimental methods is titin, which plays an important role  
in the elasticity of muscles \cite{titin}. 
The interesting segment of titin is the I-band, which is made of 
similar modules (also called repeats or domains) with an immunoglobulin-like (Ig) structure \cite{pastore,titin}. 
A single folded module can resist against pulling below a threshold. If the force exceeds the threshold, 
the domain will unfold. The unfolded domain will refold into its native state if the force is removed \cite{gaub}. 

If we pull a chain of folded Ig domains slowly, 
the necessary force to keep the protein extended will increase until one of the modules unfolds. 
This is called an unfolding event, and after that we can keep the protein extended with a much weaker force. 
As we increase the extension, the modules unfold one by one until we get a fully extended chain. 
If we plot force versus extention during the stretching, we will see a saw-tooth pattern  
(Fig. \ref{fig:saw}) \cite{gaub}. 
The stretching of titin also has been studied by different theoretical methods 
like molecular dynamic simulation \cite{karplus_titin,hoang_titin}, lattice models \cite{thirum_titin} and the thick chain (TC) model \cite{toan}.

\subsection{Models}


Many theoretical models have been developed to study macromolecules and proteins. 
Among them,  Freely Jointed Chain (FJC), Freely Rotating Chain (FRC) and Worm Like Chain (WLC) are the most famous\cite{FJC, FRC, WLC}. 
They are used intensively to model the backbone of proteins by considering the protein as a chain of only  $C_\alpha$ atoms,  and usually using an effective interaction among $C_\alpha$ atoms to recover the neglected details.  
We argue here that these models are not accurate enough to model protein backbone, and they usually   
have at least one free parameter in addition to the chain length $N$ (for example, the persistence length) 
that should be calculated by fitting of theories to experiments. Instead of using such a parameter, we can simply use the well known geometric details of peptides. 


Worm Like Chain is used to study the elasticity of macromolecules \cite{WLC}.  
It is simple and works fine whenever we can apply continuum approximation.  
It is good especially for very large and stiff macromolecules like DNA \cite{marko,rosa_WLC},  
whereas it becomes physically unjustifiable when applied to 
more coarse-grained and more flexible chains like proteins.   
For instance, in stretching of titin's Ig domain, the persistence length of the related WLC is $\sim 0.4$. 
Although this value is about the size of a single peptide unit\cite{gaub},  
WLC is frequently used in literature to fit and explain experimental data because a better alternative had  been absent.
 
Freely Jointed Chain is a chain of monomers with fixed bond length \cite{FJC}. 
In this model,  bonds' angles can have any arbitrary value, which is not the case in real proteins \cite{rama}. However, Freely Jointed Chain has an advantage over more detailed models as it has an exact solution in the case of stretching.  

Freely Rotating Chain has one more constraint than the FJC. The angles are rigid and do not change  \cite{FRC}. 
Freely Rotating Chain behaves more like real proteins compared to FJC. 
This model has been used  to study different aspect of proteins and polymers including our problem of interest \cite{netz_FRC}. 
However, it does not cover all the features of bond angles because the 
angle between $C_\alpha-C_\alpha$ bonds is not 
completely rigid in real proteins (table \ref{table:angle}) \cite{rama}. 

In our model, we consider carbon and nitrogen atoms on the backbone ($C_\alpha , C$ and $N$) and we follow the Ramachandran picture. This can be considered as a fine-grained version of a FRC or as a coarse-grained version of a Four Bead model. Our model is the simplest model that includes all the geometric properties of the backbone of proteins, and it does not need any fitting  parameters.

In this work we are interested only in the geometric details rather than the effective interactions so we can keep the model simple and clear; therefore,  we only study the mechanical properties of proteins in the entropic regime and compare the results with experiments.


\section{Model and Simulation}

\begin{figure}

\centering
  \includegraphics[width=10cm]{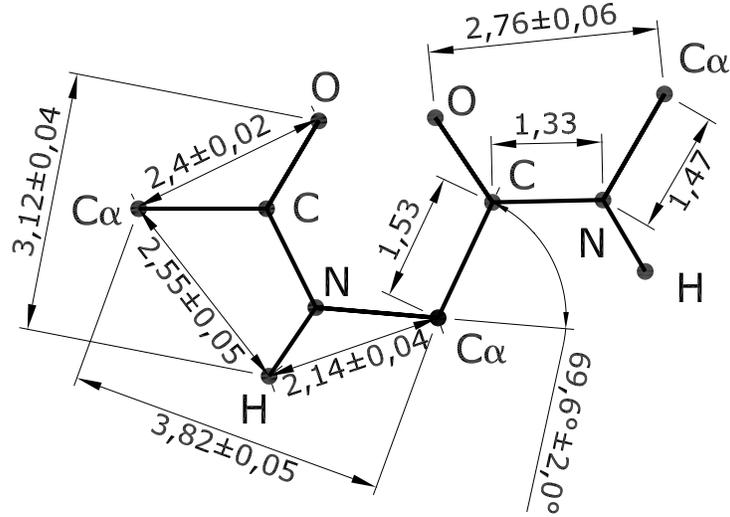}
\caption{Peptide units and their average dimensions (nanometer) and average angles, from Protein Data Bank (PDB). }
\label{fig:peptide}
\end{figure}

\begin{figure}

\centering
  \includegraphics[width=10cm]{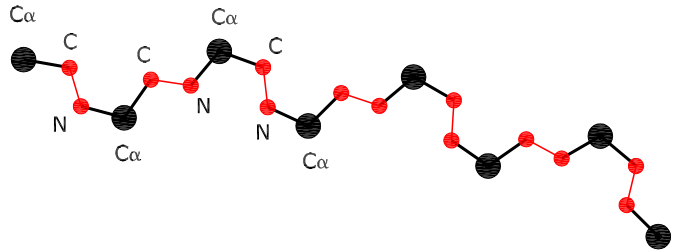}
\caption{(Color
  on-line) A Three Bead Rotating Chain. The rotations are done around $C_\alpha-C$ and $C_\alpha-N$ (bold black lines). We do not permit any rotation around $C-N$ bonds (red lines in color prints or thin gray lines that connect smaller circles in  black and white prints).}
\label{fig:chain}
\end{figure}

We use Freely Rotating Chain model 
with a small variation: some bonds can not be a pivot. 
If we show the chain as a binary sequence of 0 and 1 in which 1 represents a bond that can be a pivot 
for rotations 
and 0 as a bond that can not be a base for rotations, then our chain will be presented as:   
\begin{equation}
[101]_n \nonumber
\end{equation}
where $[ \dots ]_n$ means $n$ repeats of the argument. 
This is equivalent to considering $C$ and $N$ atoms in addition to $C_\alpha$ atoms and 
using a Ramachandran picture so that the chain only rotates around $C_\alpha-C$  and $C_\alpha-N$ bonds \cite{rama}
(Fig.  \ref{fig:chain}).  
Since the chain does not rotate around  $C-N$ bonds, 
we know the structure by only a pair of angles  ($\phi , \psi$) for each $C_\alpha$ junction. These angles  are known as Ramachandran angles \cite{rama}.  

This is the simplest model that covers Ramachandran angles which has more details than FRC, and it is simpler than the Four Bead model\cite{takada,smith,dokh}.  

To set distances and angles in our simulation, we analyzed the Protein Data Bank (PDB) 
and calculated the averages of desired quantities over all polypeptide chains in the database 
(72212 peptide units)
. The results are shown in Fig. \ref{fig:peptide} and more details are in tables \ref{table:distance} and  \ref{table:angle}. 


The largest rigid distance over the backbone of proteins is the one between two sequential $C_\alpha$ atoms; therefore, a good approximation is to consider a chain of $C_\alpha$ atoms with fixed bond length. On the other hand,  
it is not as good to assume proteins act under the Freely Rotating Chain model, because the average of angle between $C_\alpha-C_\alpha$ bond has a relatively large standard deviation and it is not completely rigid (table \ref{table:angle}).








\begin{table}
\centering
\begin{tabular}{|l|l|l|}

\hline
Atom-Atom  & Distances (nm)	& $\sigma$  \\\hline \hline
$C_{\alpha 1} C_{\alpha 2}$ &	   3.82 &	0.05 	 \\\hline 
$C_{\alpha 1}N_1$   &   2.43 &	0.05  \\\hline
$C_{\alpha 1}O_1$    &	   2.40 &	0.02  \\\hline
$C_1 C_{\alpha 2}$    &	  1.53 &		0.01  \\\hline
$C_{\alpha 1} C_1$       &1.53 &		0.01  \\\hline
$C_1 N_1$ &          1.33 &	0.04  \\\hline
$N_1 C_{\alpha 2}$ &       1.47 &	0.01  \\\hline
$N_1 C_1$   &   	   2.47  &	0.04  \\\hline
$C_1 C_2$    &   	   3.2 &	0.2  \\
\hline
\end{tabular} 

\caption{Atomic distances of the main atoms in peptides. Indexes show the order of atoms from 
left to right in the Fig. \ref{fig:peptide}. }
\label{table:distance}

\end{table}

\begin{table}
\centering
\vspace{1cm}
\begin{tabular}{|l|l|l|l|}

\hline

$\vec{A}. \vec{B}$ &	Dot products &	$\sigma$ & Angle (Degree) \\ \hline \hline
$\vec{C_{\alpha 1} C_{\alpha 2}}.\vec{C_{\alpha 2} C_{\alpha 3}} $ &	0.19 &	0.25 & 78.9  \\ \hline
$\vec{N_1 C_{\alpha 2}} . \vec{C_2 C_{\alpha 2}}$   &	0.349 &	0.032 & 69.60  \\ \hline
\end{tabular} 

\caption{Angles in peptide units. The vectors are normalized and the indexes show the order of atoms from 
left to right in Fig. \ref{fig:peptide}. You can see that FRC is not valid because 
the Standard Deviation of $\vec{C_{\alpha 1} C_{\alpha 2}}.\vec{C_{\alpha 2} C_{\alpha 3}}$ is too large compare to its  average. 
}
\label{table:angle}

\end{table}





\section{Results}

\begin{figure} 
\vspace{1cm}
\centering
\includegraphics[width=\imgwidth]{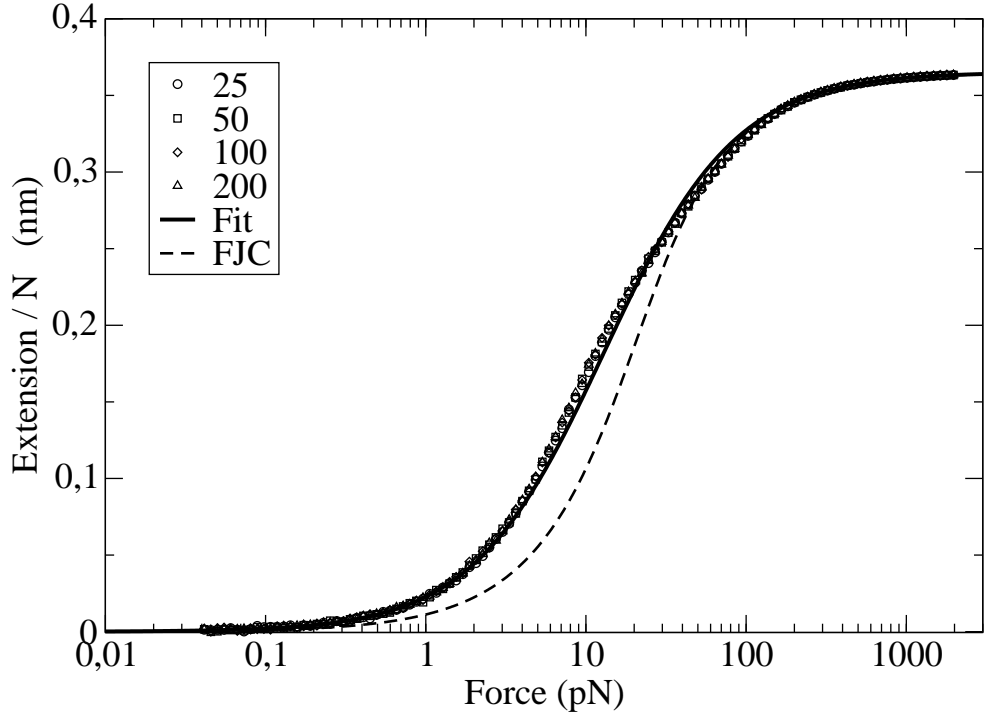}
\caption{ Results of simulations for $N= 25, 50, 100$ and $200$ peptides. All the curves have been divided by $N$. 
The curves collapse perfectly. The solid line shows Eq. \ref{equ:single} and the related parameters are in table \ref{table:saw}.}
\label{fig:simulation}
\end{figure} 

\begin{figure} 
\vspace{1cm}
\centering
\includegraphics[width=\imgwidth]{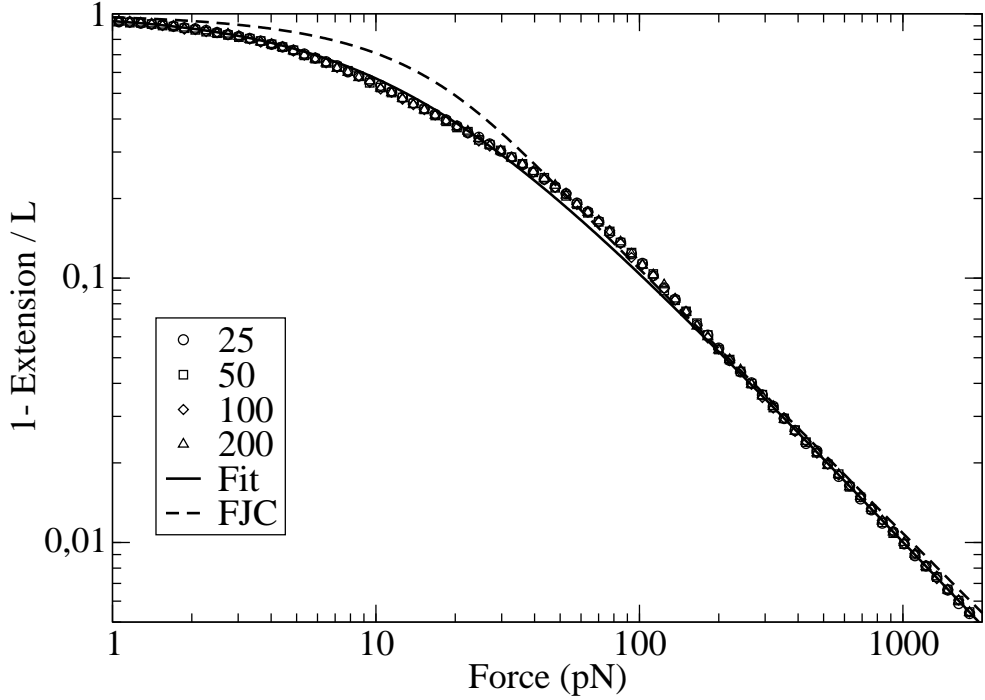}
\caption{Reduced extension versus force.}
\label{fig:1-extension}
\end{figure} 


We performed equilibrium Monte Carlo simulation and we used the standard metropolis algorithm 
 \cite{metropolis} with Pivot move \cite{pivot} and a simple Hamiltonian of the form $H=\vec{f} . \vec{r}$, where $\vec{f}$ is the pulling force and $\vec{r}$ is the position of the end of the chain while the other end is fixed at the origin. 
In the experiments, proteins are pulled slowly so we could use equilibrium Monte Carlo. 
We used the original random number generator of GNU C/C++ compiler to perform the simulation. 


We started from a force of zero and an extended structure. We performed the simulation until reaching  equilibrium, then we increased the force gradually. At each step we let the system reach the equilibrium,  
then we measured the average extension along the direction of the force. 
This process is equivalent to the AFM experiments with a soft cantilever \cite{sm}, so we can use the experimental data of this Ref. \cite{gaub}. 

We continued the process until we got a fully extended chain ($\sim 32$ nm for 89 residues). 
We repeated it for chains with different numbers of residues. We found that the extension is proportional 
to the number of residues, independent of the force. In Fig.  \ref{fig:simulation} one can see that all the extension-force curves collapse very well when their extensions have been divided by the number of residues  ($N$). 
Because of this universality, it is useful to find a suitable 
fitting function to represent results of simulation and reproduce them for later uses. The Logistic Dose-Response,  

\begin{equation}
x=\frac{a N}{1+(f_c/f)^\gamma} , 
\label{equ:single}
\end{equation}
is a good candidate. 
Although it is simple and has only three parameters, it includes all the important features of the curves:  a parameter to control the transition height ($a N$), 
another one to control the transition center ($f_c$) and one more to control the transition sharpness ($\gamma$). In Eq. \ref{equ:single} variables $x$,  $f$ and $N$ are extension, force and the number of residues, respectively. The share of each peptide units in total length is $a= 0.3640$ nm     
and  $f_c= 12.46$ pN shows the critical force which separates the swollen and the extended phases. 
The last parameter  is the fitting exponent $\gamma=1.021$ that controls the transition sharpness. If we want to fit the function to an experiment, we have to consider also the possible offset from zero.

The function \ref{equ:single} holds for the stretching of a chain without self interaction, but
we can use it to fit the saw-tooth pattern as well. To do this, we consider that the multi domain chain can have both the unfolded entropic chains and the folded domains. We know the behavior of the former from  Eq.  \ref{equ:single}, and we can approximate the latter as a rigid and inextensible objects of an unknown size. We will find its size from the peak to peak distance of the saw-tooth patterns. 
In fact, the distance between two consecutive peaks in saw-tooth patterns is equal to the length of an unfolded domain minus its end-to-end distance when it is folded.
The following

\begin{equation}
x=\frac{a n N}{1+(f_c/f)^\gamma} +  (n_{Ig}-n) R_{d} + x_0 
\label{equ:saw}
\end{equation}

\begin{table}
\centering
\begin{tabular}{|l|l|}
\hline $a$ & $0.3651 \pm 0.0004 $ nm\\ 
\hline $f_c$ & $13.0 \pm 0.1$ pN\\ 
\hline $\gamma$ & $1.072 \pm 0.004$ \\ 
\hline $R_d$ & $5.4 \pm 0.3$ nm\\
\hline 
\end{tabular} 
\caption{ Parameters of Eq. \ref{equ:saw}. Where $a$, $f_c$ and $\gamma$ has been calculated by simulation, $R_d$ has been calculated by fitting the simulation results to that of saw-tooth like patterns.}
\label{table:saw}
\end{table}

gives us the extension after the $n$th unfolding event.  Here $N$ is the number of residues in a single domain, and $n_{Ig}$ is the number of Ig repeats in the experiment. The parameter $R_{d}$ shows the distance between the first residues of two consecutive folded domains which is equal to the end to end distance of a folded Ig domain plus the length of one peptide unit. The last parameter $x_0$ is the offset from zero in experiments.
 
The first term in Eq. \ref{equ:saw} is the extension 
of unfolded domains and the second term is the contribution of folded domains to the contour length. 
It should be mentioned that we have assumed that the folded domains are rigid objects which only increase the length of the chain and do not contribute to the entropy and the force; 
therefore, we expect that the simulations fit better to the last events, when most of the domains are unfolded. 

We know $N$ and we have found $a$ and $f_c$ by simulation; therefore, only $R_{d}$ 
remains unknown and can be used as a fitting parameter to set the peak to peak distance. 
We will have a good fit if we choose $R_d$  equal to $5.4$ nm; therefore, it gives us the end to end distance of a single folded domain that is approximately $5.0$ nm. This value 
roughly agrees with the $4.3$ size taken by NMR\cite{pastore}.  
This small discrepancy ($0.7$ nm, almost twice of a peptide unit) might be due to the 
structure deformation of domains under tension. 

\begin{figure} 
\vspace{1cm}
\centering
\includegraphics[width=\imgwidth]{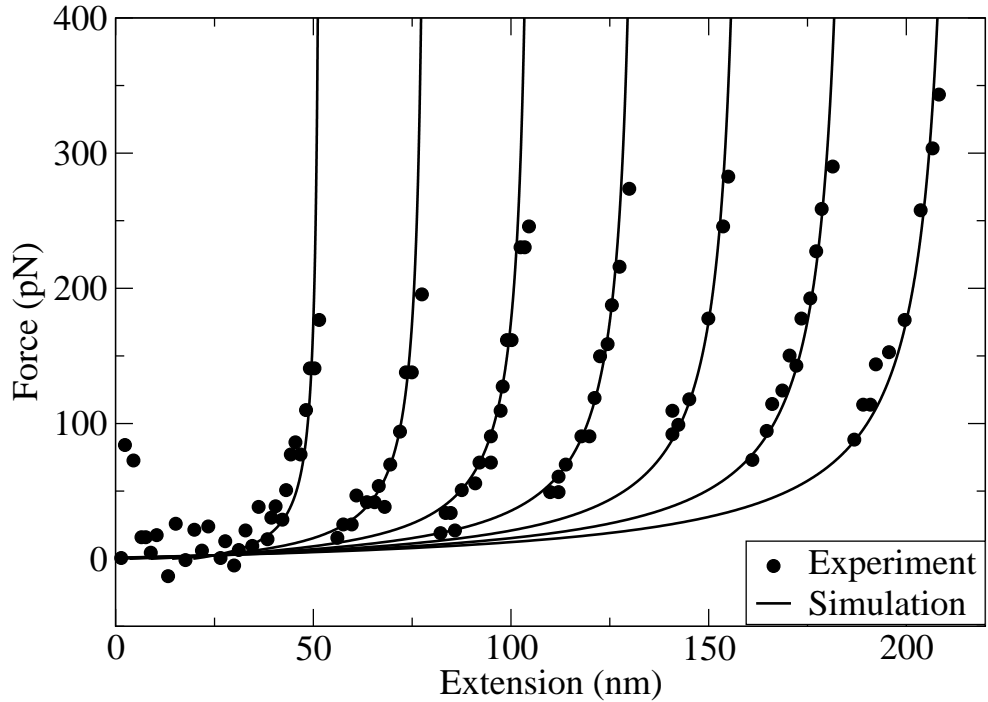}
\caption{We fitted the simulation (lines) to the experiment (circles) of \cite{gaub} by 
following the concepts of the Eq. \ref{equ:saw}. 
Related fitting parameters are in table \ref{table:saw}. }
\label{fig:saw}
\end{figure}

Figure \ref{fig:saw} displays the results of simulations and their comparisons to experimental.  The experimental data has been taken directly from the graph of 
Ref. \cite{gaub}. The parameters are displayed in table \ref{table:saw}.

\section{Conclusion}

We used a model that has only one free parameter: the number of residues. By using this model, we could fit the result of the simulation to a single unfolding event accurately. 
This shows that the primary structure of the protein is not important in the entropic regime, within the experimental precision. 

Our results suggest that this model is accurate enough to study the protein backbones in the entropic  regime. Simpler models will need to introduce fitting parameters to the model to reproduce the experimental data.

Since our model performed well in describing the mechanical properties of protein backbones in the entropic regime, one may go one step further and use it as a base for studies like \cite{amos_pre_tube} 
in which our model can reduce the number of effective interactions and present a clearer and more straightforward picture. Besides, it is easy to consider also the hydrogen bonds along the backbone, since it is sufficient to include oxygen and hydrogen atoms of the backbone.


\bibliographystyle{apsrev} 

\bibliography{prc}

\begin{thebibliography}{25}
\expandafter\ifx\csname natexlab\endcsname\relax\def\natexlab#1{#1}\fi
\expandafter\ifx\csname bibnamefont\endcsname\relax
  \def\bibnamefont#1{#1}\fi
\expandafter\ifx\csname bibfnamefont\endcsname\relax
  \def\bibfnamefont#1{#1}\fi
\expandafter\ifx\csname citenamefont\endcsname\relax
  \def\citenamefont#1{#1}\fi
\expandafter\ifx\csname url\endcsname\relax
  \def\url#1{\texttt{#1}}\fi
\expandafter\ifx\csname urlprefix\endcsname\relax\def\urlprefix{URL }\fi
\providecommand{\bibinfo}[2]{#2}
\providecommand{\eprint}[2][]{\url{#2}}

\bibitem[{\citenamefont{Rief et~al.}(1997)\citenamefont{Rief, Gautel,
  Oesterhelt, Fernandez, and Gaub}}]{gaub}
\bibinfo{author}{\bibfnamefont{M.}~\bibnamefont{Rief}},
  \bibinfo{author}{\bibfnamefont{M.}~\bibnamefont{Gautel}},
  \bibinfo{author}{\bibfnamefont{F.}~\bibnamefont{Oesterhelt}},
  \bibinfo{author}{\bibfnamefont{J.~M.} \bibnamefont{Fernandez}},
  \bibnamefont{and} \bibinfo{author}{\bibfnamefont{H.~E.} \bibnamefont{Gaub}},
  \bibinfo{journal}{Science} \textbf{\bibinfo{volume}{276}},
  \bibinfo{pages}{1109} (\bibinfo{year}{1997}).

\bibitem[{\citenamefont{Z. et~al.}(1997)\citenamefont{Z., Kellermayer, Smith,
  Granzier, and Bustamante}}]{optic1}
\bibinfo{author}{\bibfnamefont{M.~S.} \bibnamefont{Z.}},
  \bibinfo{author}{\bibnamefont{Kellermayer}},
  \bibinfo{author}{\bibfnamefont{S.~B.} \bibnamefont{Smith}},
  \bibinfo{author}{\bibfnamefont{H.~L.} \bibnamefont{Granzier}},
  \bibnamefont{and}
  \bibinfo{author}{\bibfnamefont{C.}~\bibnamefont{Bustamante}},
  \bibinfo{journal}{Science} \textbf{\bibinfo{volume}{276}},
  \bibinfo{pages}{1112} (\bibinfo{year}{1997}).

\bibitem[{\citenamefont{Tskhovrebova et~al.}(1997)\citenamefont{Tskhovrebova,
  Trinick, Sleep, and Simmons}}]{optic2}
\bibinfo{author}{\bibfnamefont{L.}~\bibnamefont{Tskhovrebova}},
  \bibinfo{author}{\bibfnamefont{J.}~\bibnamefont{Trinick}},
  \bibinfo{author}{\bibfnamefont{J.~A.} \bibnamefont{Sleep}}, \bibnamefont{and}
  \bibinfo{author}{\bibfnamefont{R.~M.} \bibnamefont{Simmons}},
  \bibinfo{journal}{Nature} \textbf{\bibinfo{volume}{387}}, \bibinfo{pages}{308
  } (\bibinfo{year}{1997}).

\bibitem[{\citenamefont{Best et~al.}(2001)\citenamefont{Best, Li, Steward,
  Daggett, and Clarke}}]{non_machanical}
\bibinfo{author}{\bibfnamefont{R.~B.} \bibnamefont{Best}},
  \bibinfo{author}{\bibfnamefont{B.}~\bibnamefont{Li}},
  \bibinfo{author}{\bibfnamefont{A.}~\bibnamefont{Steward}},
  \bibinfo{author}{\bibfnamefont{V.}~\bibnamefont{Daggett}}, \bibnamefont{and}
  \bibinfo{author}{\bibfnamefont{J.}~\bibnamefont{Clarke}},
  \bibinfo{journal}{Biophys Journal} \textbf{\bibinfo{volume}{81}},
  \bibinfo{pages}{2344} (\bibinfo{year}{2001}).

\bibitem[{\citenamefont{Fisher et~al.}(2000)\citenamefont{Fisher, Marszalek,
  and Fernandez}}]{fernandez_titin}
\bibinfo{author}{\bibfnamefont{T.~E.} \bibnamefont{Fisher}},
  \bibinfo{author}{\bibfnamefont{P.~E.} \bibnamefont{Marszalek}},
  \bibnamefont{and} \bibinfo{author}{\bibfnamefont{J.~M.}
  \bibnamefont{Fernandez}}, \bibinfo{journal}{Nature structural biology}
  \textbf{\bibinfo{volume}{7}}, \bibinfo{pages}{719 } (\bibinfo{year}{2000}).

\bibitem[{\citenamefont{Labeit and Kolmerer}(1995)}]{titin}
\bibinfo{author}{\bibfnamefont{S.}~\bibnamefont{Labeit}} \bibnamefont{and}
  \bibinfo{author}{\bibfnamefont{B.}~\bibnamefont{Kolmerer}},
  \bibinfo{journal}{Science} \textbf{\bibinfo{volume}{270}},
  \bibinfo{pages}{293} (\bibinfo{year}{1995}).

\bibitem[{\citenamefont{Improta et~al.}(1996)\citenamefont{Improta, Politou,
  and Pastore}}]{pastore}
\bibinfo{author}{\bibfnamefont{S.}~\bibnamefont{Improta}},
  \bibinfo{author}{\bibfnamefont{A.}~\bibnamefont{Politou}}, \bibnamefont{and}
  \bibinfo{author}{\bibfnamefont{A.}~\bibnamefont{Pastore}},
  \bibinfo{journal}{Structure} \textbf{\bibinfo{volume}{4}},
  \bibinfo{pages}{323} (\bibinfo{year}{1996}).

\bibitem[{\citenamefont{Paci and Karplus}(2000)}]{karplus_titin}
\bibinfo{author}{\bibfnamefont{E.}~\bibnamefont{Paci}} \bibnamefont{and}
  \bibinfo{author}{\bibfnamefont{M.}~\bibnamefont{Karplus}},
  \bibinfo{journal}{Proc Natl Acad Sci USA} \textbf{\bibinfo{volume}{97}},
  \bibinfo{pages}{6521} (\bibinfo{year}{2000}).

\bibitem[{\citenamefont{Cieplak et~al.}(2004)\citenamefont{Cieplak, Hoang, and
  Robbins}}]{hoang_titin}
\bibinfo{author}{\bibfnamefont{M.}~\bibnamefont{Cieplak}},
  \bibinfo{author}{\bibfnamefont{T.~X.} \bibnamefont{Hoang}}, \bibnamefont{and}
  \bibinfo{author}{\bibfnamefont{M.~O.} \bibnamefont{Robbins}},
  \bibinfo{journal}{Proteins} \textbf{\bibinfo{volume}{56}},
  \bibinfo{pages}{285} (\bibinfo{year}{2004}).

\bibitem[{\citenamefont{Klimov and Thirumalai}(1999)}]{thirum_titin}
\bibinfo{author}{\bibfnamefont{D.~K.} \bibnamefont{Klimov}} \bibnamefont{and}
  \bibinfo{author}{\bibfnamefont{D.}~\bibnamefont{Thirumalai}},
  \bibinfo{journal}{Proc Natl Acad Sci USA} \textbf{\bibinfo{volume}{96}},
  \bibinfo{pages}{6166} (\bibinfo{year}{1999}).

\bibitem[{\citenamefont{Toan et~al.}(2005)\citenamefont{Toan, Marenduzzo, and
  Micheletti}}]{toan}
\bibinfo{author}{\bibfnamefont{N.~M.} \bibnamefont{Toan}},
  \bibinfo{author}{\bibfnamefont{D.}~\bibnamefont{Marenduzzo}},
  \bibnamefont{and}
  \bibinfo{author}{\bibfnamefont{C.}~\bibnamefont{Micheletti}},
  \bibinfo{journal}{Biophysical Journal} \textbf{\bibinfo{volume}{89}},
  \bibinfo{pages}{80} (\bibinfo{year}{2005}).

\bibitem[{\citenamefont{Kramers}(1946)}]{FJC}
\bibinfo{author}{\bibfnamefont{H.}~\bibnamefont{Kramers}},
  \bibinfo{journal}{Journal of chemical physics} \textbf{\bibinfo{volume}{14}},
  \bibinfo{pages}{415} (\bibinfo{year}{1946}).

\bibitem[{\citenamefont{Flory}(1969)}]{FRC}
\bibinfo{author}{\bibfnamefont{P.~J.} \bibnamefont{Flory}},
  \emph{\bibinfo{title}{Statistical Mechaics of Chain Molecules}}
  (\bibinfo{publisher}{Wiley}, \bibinfo{address}{New York},
  \bibinfo{year}{1969}).

\bibitem[{\citenamefont{Kratky and Porod}(1949)}]{WLC}
\bibinfo{author}{\bibfnamefont{O.}~\bibnamefont{Kratky}} \bibnamefont{and}
  \bibinfo{author}{\bibfnamefont{G.}~\bibnamefont{Porod}},
  \bibinfo{journal}{Rec. Trav. Chim.} \textbf{\bibinfo{volume}{68}},
  \bibinfo{pages}{1106} (\bibinfo{year}{1949}).

\bibitem[{\citenamefont{Marko and Siggia}(1995)}]{marko}
\bibinfo{author}{\bibfnamefont{J.}~\bibnamefont{Marko}} \bibnamefont{and}
  \bibinfo{author}{\bibfnamefont{E.}~\bibnamefont{Siggia}},
  \bibinfo{journal}{Macromolecules} pp. \bibinfo{pages}{8759--8770}
  (\bibinfo{year}{1995}).

\bibitem[{\citenamefont{Rosa et~al.}(2003)\citenamefont{Rosa, Hoang,
  Marenduzzo, and Maritan}}]{rosa_WLC}
\bibinfo{author}{\bibfnamefont{A.}~\bibnamefont{Rosa}},
  \bibinfo{author}{\bibfnamefont{T.}~\bibnamefont{Hoang}},
  \bibinfo{author}{\bibfnamefont{D.}~\bibnamefont{Marenduzzo}},
  \bibnamefont{and} \bibinfo{author}{\bibfnamefont{A.}~\bibnamefont{Maritan}},
  \bibinfo{journal}{Macromolecules} \textbf{\bibinfo{volume}{36}},
  \bibinfo{pages}{10095} (\bibinfo{year}{2003}).

\bibitem[{\citenamefont{Ramachandran and Sasisekharan}(1968)}]{rama}
\bibinfo{author}{\bibfnamefont{G.~N.} \bibnamefont{Ramachandran}}
  \bibnamefont{and}
  \bibinfo{author}{\bibfnamefont{V.}~\bibnamefont{Sasisekharan}},
  \bibinfo{journal}{advanced in protein chemistry}
  \textbf{\bibinfo{volume}{23}}, \bibinfo{pages}{283} (\bibinfo{year}{1968}).

\bibitem[{\citenamefont{Livadaru et~al.}(2003)\citenamefont{Livadaru, Netz, and
  Kreuzer}}]{netz_FRC}
\bibinfo{author}{\bibfnamefont{L.}~\bibnamefont{Livadaru}},
  \bibinfo{author}{\bibfnamefont{R.~R.} \bibnamefont{Netz}}, \bibnamefont{and}
  \bibinfo{author}{\bibfnamefont{H.~J.} \bibnamefont{Kreuzer}},
  \bibinfo{journal}{Macromolecules} \textbf{\bibinfo{volume}{36}},
  \bibinfo{pages}{3732} (\bibinfo{year}{2003}).

\bibitem[{\citenamefont{{Takada} et~al.}(1999)\citenamefont{{Takada},
  {Luthey-Schulten}, and {Wolynes}}}]{takada}
\bibinfo{author}{\bibfnamefont{S.}~\bibnamefont{{Takada}}},
  \bibinfo{author}{\bibfnamefont{Z.}~\bibnamefont{{Luthey-Schulten}}},
  \bibnamefont{and} \bibinfo{author}{\bibfnamefont{P.~G.}
  \bibnamefont{{Wolynes}}}, \bibinfo{journal}{\jcp}
  \textbf{\bibinfo{volume}{110}}, \bibinfo{pages}{11616}
  (\bibinfo{year}{1999}).

\bibitem[{\citenamefont{Smith and Hall}(2001)}]{smith}
\bibinfo{author}{\bibfnamefont{A.~V.} \bibnamefont{Smith}} \bibnamefont{and}
  \bibinfo{author}{\bibfnamefont{C.}~\bibnamefont{Hall}},
  \bibinfo{journal}{Proteins Structure Function and Genetics}
  \textbf{\bibinfo{volume}{44}}, \bibinfo{pages}{344} (\bibinfo{year}{2001}).

\bibitem[{\citenamefont{Ding et~al.}(2003)\citenamefont{Ding, Borreguero,
  Buldyrey, Stanley, and Dokholyan}}]{dokh}
\bibinfo{author}{\bibfnamefont{F.}~\bibnamefont{Ding}},
  \bibinfo{author}{\bibfnamefont{J.}~\bibnamefont{Borreguero}},
  \bibinfo{author}{\bibfnamefont{S.}~\bibnamefont{Buldyrey}},
  \bibinfo{author}{\bibfnamefont{H.}~\bibnamefont{Stanley}}, \bibnamefont{and}
  \bibinfo{author}{\bibfnamefont{N.}~\bibnamefont{Dokholyan}},
  \bibinfo{journal}{Proteins Structure Function and Genetics}
  \textbf{\bibinfo{volume}{53}}, \bibinfo{pages}{220} (\bibinfo{year}{2003}).

\bibitem[{\citenamefont{Metropolis}(1953)}]{metropolis}
\bibinfo{author}{\bibfnamefont{N.}~\bibnamefont{Metropolis}},
  \bibinfo{journal}{Journal ofChemical Physics} \textbf{\bibinfo{volume}{21}},
  \bibinfo{pages}{1087} (\bibinfo{year}{1953}).

\bibitem[{\citenamefont{Madras and Sokal}(1988)}]{pivot}
\bibinfo{author}{\bibfnamefont{N.}~\bibnamefont{Madras}} \bibnamefont{and}
  \bibinfo{author}{\bibfnamefont{A.}~\bibnamefont{Sokal}},
  \bibinfo{journal}{Journal of Statistical Physics}
  \textbf{\bibinfo{volume}{50}}, \bibinfo{pages}{109} (\bibinfo{year}{1988}).

\bibitem[{\citenamefont{Kreuzer et~al.}(2001)\citenamefont{Kreuzer, Payne, and
  Livadaru}}]{sm}
\bibinfo{author}{\bibfnamefont{H.~J.} \bibnamefont{Kreuzer}},
  \bibinfo{author}{\bibfnamefont{S.~H.} \bibnamefont{Payne}}, \bibnamefont{and}
  \bibinfo{author}{\bibfnamefont{L.}~\bibnamefont{Livadaru}},
  \bibinfo{journal}{Biophysical Journal} \textbf{\bibinfo{volume}{80}},
  \bibinfo{pages}{2505} (\bibinfo{year}{2001}).

\bibitem[{\citenamefont{Banavar et~al.}(2004)\citenamefont{Banavar, Hoang,
  Maritan, Seno, and Trovato}}]{amos_pre_tube}
\bibinfo{author}{\bibfnamefont{J.~R.} \bibnamefont{Banavar}},
  \bibinfo{author}{\bibfnamefont{T.~X.} \bibnamefont{Hoang}},
  \bibinfo{author}{\bibfnamefont{A.}~\bibnamefont{Maritan}},
  \bibinfo{author}{\bibfnamefont{F.}~\bibnamefont{Seno}}, \bibnamefont{and}
  \bibinfo{author}{\bibfnamefont{A.}~\bibnamefont{Trovato}},
  \bibinfo{journal}{Physical Review E} \textbf{\bibinfo{volume}{70}}
  (\bibinfo{year}{2004}).

\end{thebibliography}

\end{document}